\documentclass[12pt]{iopart}
\usepackage{epsfig} 
\begin{document}
\title[PHENIX event-by-event average $p_T$ fluctuations]{Event-by-event average $p_T$ fluctuations in 
$\sqrt{s_{NN}}=200$~GeV Au+Au and p+p collisions in PHENIX: measurements and jet contribution simulations}

\author{M J Tannenbaum\dag \  for the PHENIX Collaboration\footnote[2]{For the full PHENIX Collaboration author list and
acknowledgments, see Appendix "Collaborations" of this volume.}}

\address{\dag\ Physics Dept., 510c, Brookhaven National Laboratory, Upton, NY 11973-5000 USA}

\begin{abstract}
 Small, but significant non-random fluctuations in event-by-event average $p_T$ have been observed in Au+Au collisions at $\sqrt{s_{NN}}=200$ GeV by the PHENIX collaboration. These are consistent with being caused by correlations due to jets at large $p_T$, where the measured suppression must be included to reproduce the centrality dependence of the non-random fluctuations.  
\end{abstract}


\section{The event-by-event average $p_T$ distribution is not a Gaussian, it's a Gamma distribution.}

      The single particle inclusive $p_T$ distribution averaged over all particles in all events in a p-p experiment (inclusive) or in a given centrality class in an A+A experiment (semi-inclusive) is usually written in the form: 
\begin{equation}
{d\sigma\over {p_T dp_T}}=b^2 e^{-b p_T} \qquad\mbox{or}\qquad
{d\sigma\over {dp_T}}=b^2 p_T e^{-b p_T} \label{eq:ptdist}
\qquad .
\end{equation}
Equation~\ref{eq:ptdist} represents a Gamma 
distribution with $p=2$, where, $\langle p_T\rangle=p/b$, $\sigma_{p_T}/\langle p_T\rangle = 1/\sqrt{p}$ and typically $b=6$ (GeV/c)$^{-1}$ for p-p collisions. The `inverse slope 
parameter' $T=1/b$ is sometimes referred to as the `Temperature 
parameter'.  

\begin{figure}[htb]
\vspace*{-0.12in}\begin{center}
\hspace*{-0.2in}

\includegraphics[scale=1.0,angle=0,height=2in, width=5.0in]{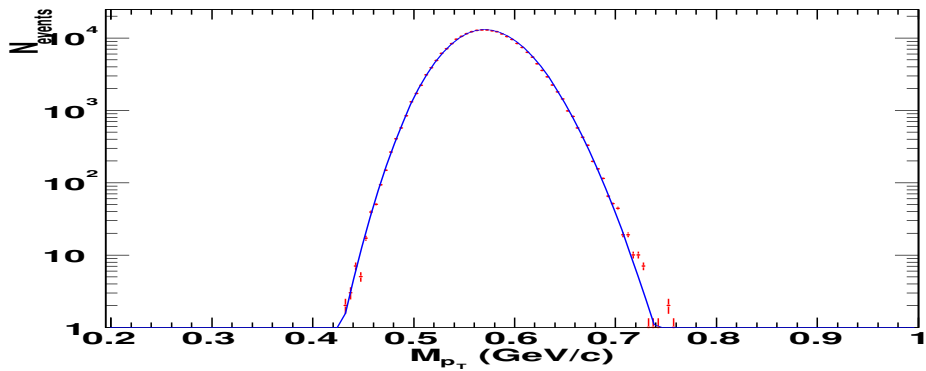}

\end{center}\vspace*{-0.25in}
\caption[]{PHENIX mixed-event distribution for the 0-5\% centrality class (data points) compared to  equation~\ref{eq:nbdgamma} (curve).  \label{fig:mixedevtcheck}}

\end{figure}

 For events with $n$ detected charged particles with magnitudes of transverse momenta, $p_{T_i}$, the event-by-event average $p_T$, denoted $M_{p_T}$ is defined as: 

 \begin{equation}
M_{p_T}=\overline{p_T}={1\over n} \sum_{i=1}^n p_{T_i}={1\over 
n} E_{Tc} \qquad .\label{eq:defMpT}
\end{equation}
For the case of statistical independent emission, where the fluctuations are purely random, an analytical formula for the distribution in $M_{p_T}$ can be 
obtained assuming negative binomial (NBD) distributed 
event-by-event multiplicity, with Gamma distributed semi-inclusive 
$p_T$ spectra.~\cite{elsevavgpt} The formula depends on the 4 semi-inclusive parameters $\langle 
n\rangle$, 
$1/k$, $b$ and $p$ which are derived from the means and 
standard deviations of the semi-inclusive $p_T$ and multiplicity 
distributions, $\langle 
n\rangle$, $\sigma_n$, $\langle p_T\rangle$, $\sigma_{p_T}$: 
\begin{equation}
       f(y)=
\sum_{n={n_{\rm min}}}^{n_{\rm max}} f_{\rm NBD}(n,1/k,\langle 
n\rangle) 
\, f_{\Gamma}(y,np,nb) \qquad ,   
\label{eq:nbdgamma}
\end{equation} 
where $y=M_{p_T}$. For fixed $n$, and purely random fluctuations, the mean and standard deviation of $M_{p_T}$ follow the expected behavior, $\langle M_{p_T}\rangle=\langle p_T\rangle$, $\sigma_{M_{p_T}}=\sigma_{p_T}/\sqrt{n}$.  In PHENIX, equation~\ref{eq:nbdgamma} is used to confirm the randomness of mixed-events (Figure~\ref{fig:mixedevtcheck}).
\vspace*{-0.24in}

\section{Measurement of non-random fluctuations in PHENIX}
\vspace*{-0.05in}
Mixed-events are used to define the baseline for random fluctuations of $M_{p_T}$ in  PHENIX~\cite{PRC66,PX200}. This has the advantage of effectively removing any residual detector-dependent effects. The event-by-event average distributions
are very sensitive to the number of tracks in the event (denoted $n$ or $N_{tracks}$), so the mixed event sample is produced with the {\em identical} $N_{tracks}$ distribution as the data. Additionally, no two tracks from the same data event are
placed in the same mixed event in order to remove any intra-event correlations in $p_T$. Finally, $\langle M_{p_T}\rangle$ must exactly match the semi-inclusive $\langle p_T\rangle$.
As noted above, the randomness of $M_{p_T}$ for the mixed-event sample is tested by comparison to equation~\ref{eq:nbdgamma}. 
Figure~\ref{fig:mixedevtcheck} shows the excellent agreement
between the calculation and the mixed event $M_{p_T}$ distributions for the 0-5\% centrality class. The standard deviations, $\sigma_{M_{p_T}}$, differ by less than 0.04\%. This represents the maximum error from any effects introduced by the event mixing procedure.

\begin{figure}[htb]
\begin{center}
\begin{tabular}{cc}

\includegraphics[scale=0.9,angle=0]{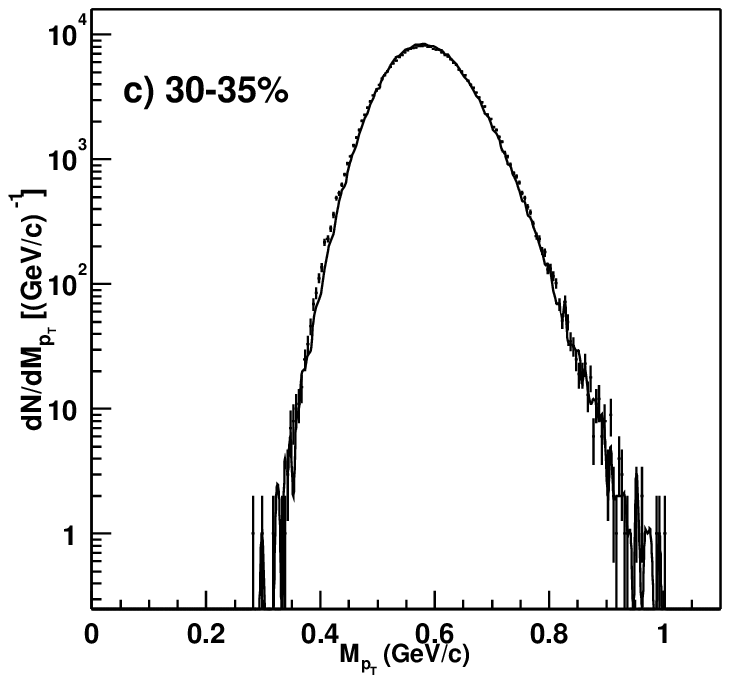}&
\includegraphics[scale=0.9,angle=0]{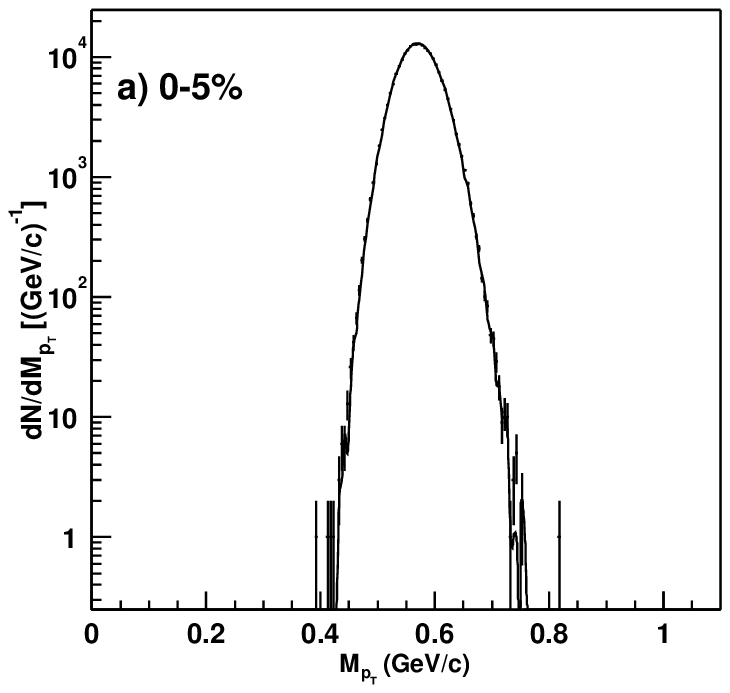}

\end{tabular}
\end{center}\vspace*{-0.25in}
\caption[]{$M_{p_T}$ for 30-35\% and 0-5\% centrality classes: data (points) mixed-events (histogram). \label{fig:MpT}}

\end{figure}

\begin{figure}[htb]
\begin{center}

\includegraphics[scale=0.60,angle=0]{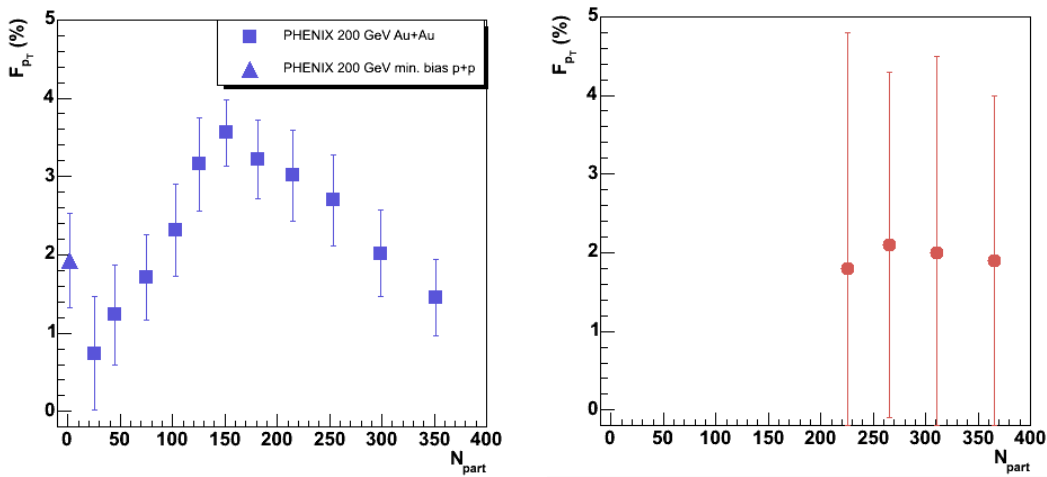}
\end{center}\vspace*{-0.25in}
\caption[]{$F_{p_T}$ in \% :(left) Au+Au, p+p, $\sqrt{s_{NN}}=200$ GeV, (right) Au+Au 130 GeV \label{fig:newvsold}}

\end{figure}

        The measured $M_{p_T}$ distributions for the data in two centrality classes for $\sqrt{s_{NN}}=200$ GeV Au+Au collisions in PHENIX are shown in Figure~\ref{fig:MpT} (data points) compared to the mixed-event distributions (histograms).   
The non-Gaussian, Gamma distribution shape of the $M_{p_T}$ distributions is evident.
The difference between the data and the mixed-event random baseline distributions is barely visible to the naked eye. The non-random fluctuation is quantified by the percent difference of the standard deviations of $M_{p_T}$ for the data and the mixed-event (random) sample:
	\begin{equation}
	F_{p_T}\equiv \frac{\sigma_{M_{p_T},{\rm data}}-\sigma_{M_{p_T},{\rm mixed}}}{\sigma_{M_{p_T},{\rm mixed}}} \qquad .
	\label{eq:def:FpT}
	\end{equation}
The results are shown as a function of centrality represented by $N_{part}$ in Figure~\ref{fig:newvsold} compared to the previous PHENIX measurement at $\sqrt{s_{NN}}=130$ GeV.~\cite{PRC66} 
The errors shown are systematic errors due to time-dependent detector variations. Comparatively,  statistical errors are negligible. The systematic error is calculated from the rms variation of $F_{p_T}$ from 10 independent subsets of the data. The improvement over the $\sqrt{s_{NN}}=130$ GeV data is due to 3 times larger solid angle (larger $N_{tracks}$), better tracking and more statistics.~\cite{PX200}

\begin{figure}[!thb]
\begin{center}
\begin{tabular}{cc}

\includegraphics[scale=0.4,angle=0]{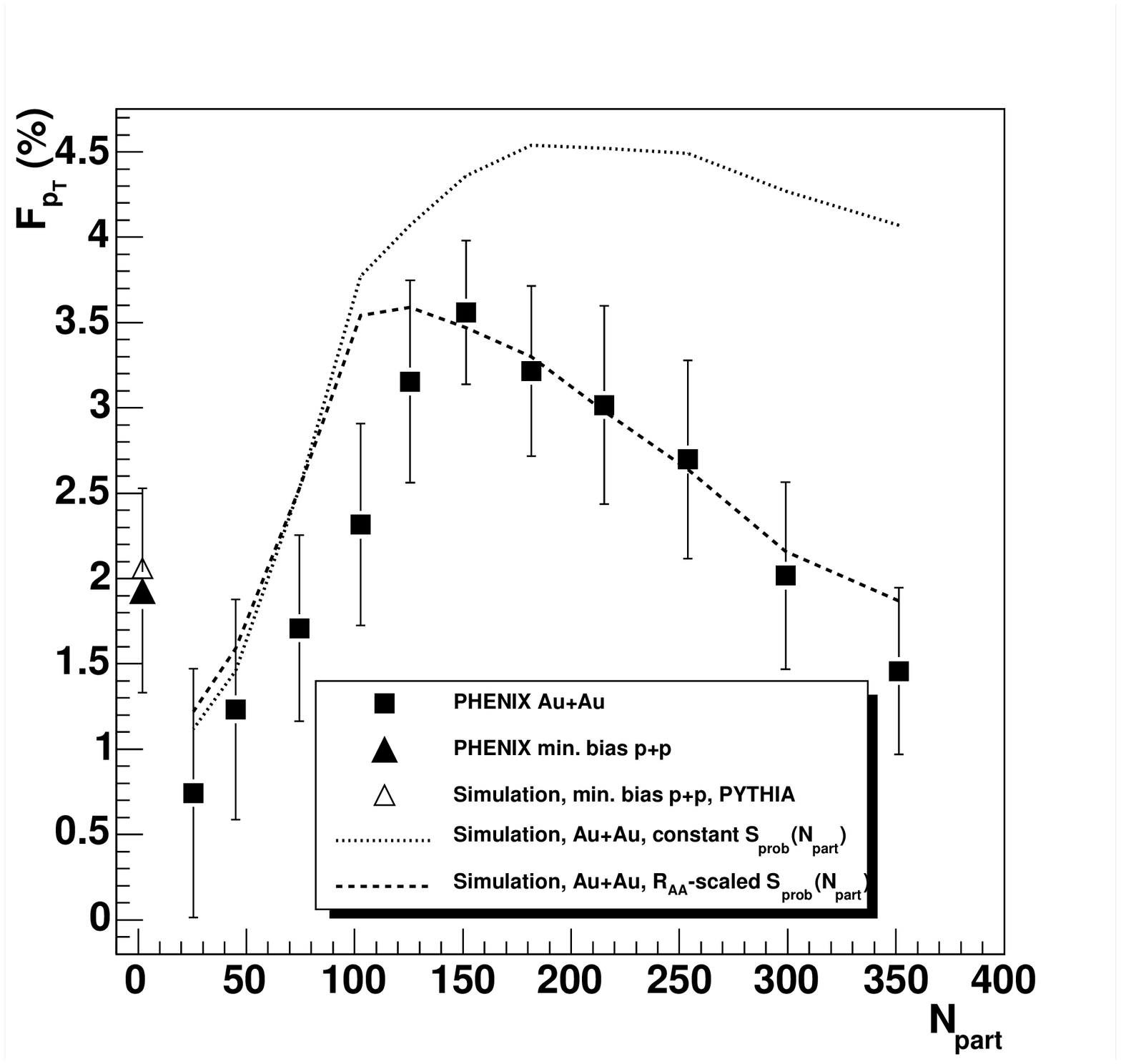}&
\includegraphics[scale=0.4,angle=0]{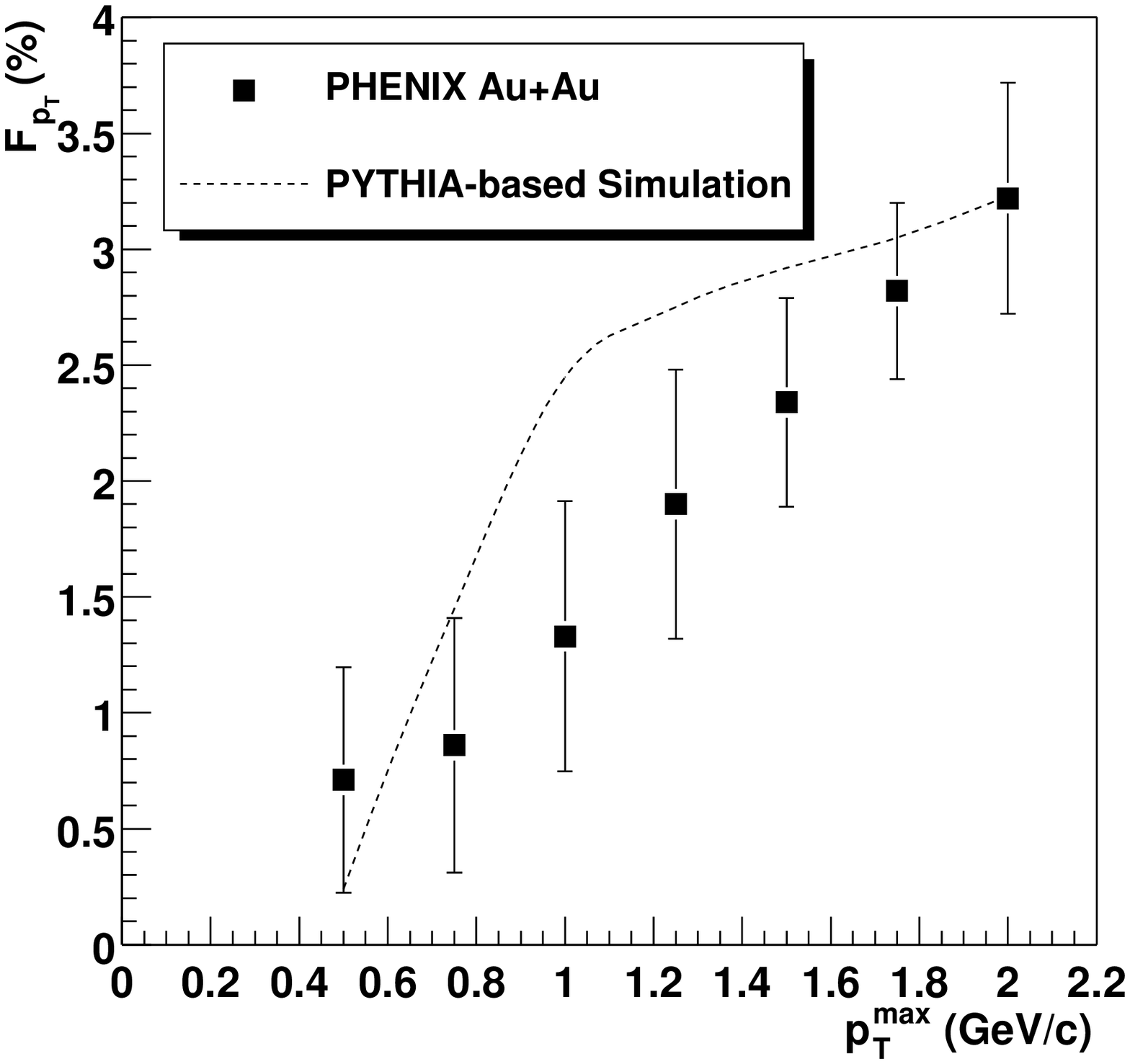}

\end{tabular}
\end{center}\vspace*{-0.25in}
\caption{$F_{p_T}$ vs centrality and $p_{T}^{\rm max}$ compared to simulations.  \label{fig:Ft}}

\end{figure}

	The dependence of $F_{p_T}$ on $N_{part}$ is striking. To further understand this dependence and the source of these non-random fluctuations, $F_{p_T}$ was measured over a varying $p_T$ range, \mbox{0.2 GeV/c $\leq p_T\leq p_T^{\rm max}$} (Figure~\ref{fig:Ft}), where $p_T^{\rm max}=2.0$ GeV/c for the $N_{part}$ dependence.   The increase of $F_{p_T}$ with $p_T^{\rm max}$ suggests elliptic flow or jet origin. This was investigated using a Monte Carlo simulation of correlations due to elliptic flow and jets in the PHENIX acceptance. The flow was significant only in the lowest centrality bin and negligible ($F_{p_T} <0.1$\%) at higher centralities. Jets were simulated by embedding [at a uniform rate per generated particle, $S_{prob}(N_{part})$] p-p hard-scattering events from the PYTHIA event generator into simulated Au+Au events assembled at random according to the measured $N_{tracks}$ and semi-inclusive $p_T$ distributions. This changed $\langle p_T\rangle$ and $\sigma_{p_T}$ by less than 0.1\% . $S_{prob}(N_{part})$ was either constant for all centrality classes, or scaled by the measured hard-scattering suppression factor $R_{AA}(N_{part})$ for $p_T > 4.5$ GeV/c.~\cite{PXRAA} A value $F_{p_T}=2.06$\% for p-p collisions was extracted from pure PYTHIA events in the PHENIX acceptance in agreement with the p-p measurement. The value of $S_{prob}(N_{part})$ was chosen so that the simulation with $S_{prob}(N_{part})\times R_{AA}(N_{part})$ agreed with the data at $N_{part}=182$. The centrality and $p_T^{\rm max}$ dependences of the measured $F_{p_T}$ match the simulation very well, but only when the $R_{AA}$ scaling is included. 

	A less experiment-dependent method to compare non-random fluctuations is to assume that the entire $F_{p_T}$ is due to temperature fluctuations of the initial state, with rms variation $\sigma_{T}/\langle T\rangle$:~\cite{stan,PRC66}
	\begin{equation}
	F_{p_T}=\frac{(\langle n\rangle -1)}{2}\frac{\sigma^{2}_{T}/\langle T\rangle^2}{\sigma^{2}_{p_T}/\langle p_T\rangle^2}={p\over 2}{(\langle n\rangle -1)} \frac{\sigma^{2}_{T}}{\langle T\rangle^2} \qquad .
	\label{eq:sigmaT}
	\end{equation}
This yields $\sigma_{T}/\langle T\rangle$=1.8\% for central collisions and 3.7\% at the peak of $F_{p_T}$, which puts severely small limits on the critical-fluctuations that were expected. 

\end{document}